\RequirePackage{fix-cm}
\documentclass[twocolumn,epjc3]{svjour3}  
\smartqed  % flush right qed marks, e.g. at end of proof
\RequirePackage{graphicx}
\RequirePackage{caption}
\RequirePackage{subcaption}
\RequirePackage[separate-uncertainty=true,per-mode=symbol]{siunitx}
\RequirePackage{isotope}
\RequirePackage{orcidlink}
\RequirePackage{lineno}
\RequirePackage{amsmath}

\newcommand\CoolingPower{\qty{118\pm3}{\watt}}
\newcommand\HeatingPower{\qty{121\pm3}{\watt}}
\newcommand\ElectricalPowerThreeBar{\qty{386\pm1}{\watt}}
\newcommand\ElectricalPowerFourBar{\qty{393\pm1}{\watt}}
\newcommand\HeatLossesThreeBar{\qty{15.1\pm0.5}{\watt}}
\newcommand\HeatLossesFourBar{\qty{14.2\pm0.5}{\watt}}

\newcommand\COPDesignIdealFourBar{2.3}

\newcommand\COPDesignIdealThreeBar{3.0}
\newcommand\COPMeasuredRealThreeBar{$0.30\pm0.01$} % no mistake are identical with 4.3 bar
\newcommand\COPMeasuredDistThreeBar{$0.60\pm0.02$}

\newcommand\VirtualPurificationFlow{\qty{3.1}{\kilo \gram \per \hour}}

\newcommand\MaxCompressorFlowFourBar{\qty{17.9\pm0.2}{slpm}}
\newcommand\MZeroFlowFourBar{\qty{1.8\pm0.1}{slpm}}
\newcommand\MFullWattFlowFourBar{\qty{15.4\pm0.4}{slpm}}

\newcommand\MaxCompressorFlowThreeBar{\qty{18.4\pm0.2}{slpm}}
\newcommand\MZeroFlowThreeBar{\qty{1.9\pm0.1}{slpm}}
\newcommand\MFullWattFlowThreeBar{\qty{14.9\pm0.4}{slpm}}

\newcommand\EtaThermal{\qty{0.89\pm0.03}{}}
\newcommand\EtaBypass{\qty{0.92\pm0.01}{}}
\newcommand\EtaCompressor{\qty{0.18\pm0.01}{}}
\newcommand\EtaPressure{\qty{0.76\pm0.01}{}}
\newcommand\EtaHeatExchanger{\qty{0.87\pm0.01}{}}

\usepackage{booktabs}

\journalname{Eur. Phys. J. C}
\begin{document}

\title{Proof-of-concept of a xenon-based cryogenic heat pump demonstrator for future liquid xenon observatories}

\author{P. Schulte\orcidlink{0009-0008-9029-3092}\thanksref{addr1,email1}
        \and
        D. Wenz\orcidlink{0009-0004-5242-3571}\thanksref{addr1,email2}
        \and
        {L. Alth\"user}\orcidlink{0000-0002-5468-4298}\thanksref{addr1}
        \and
        R. Braun\orcidlink{0009-0007-0706-3054}\thanksref{addr1}
        \and
        V. Hannen\orcidlink{0000-0002-2944-8373}\thanksref{addr1}
        \and
        C. Huhmann\thanksref{addr1}
        \and
        D. Koke\orcidlink{0000-0002-8887-5527}\thanksref{addr1}
        \and
        Y.-T. Lin\orcidlink{0000-0003-3631-1655}\thanksref{addr1}
        \and
        P. Unkhoff\orcidlink{0009-0001-4054-3179}\thanksref{addr1}
        \and
        C. Weinheimer\orcidlink{0000-0002-4083-9068}\thanksref{addr1}
}
\institute{Institut f\"ur Kernphysik, Universit\"at M\"unster, M\"unster, Germany \label{addr1}
}
\thankstext{email1}{\texttt{p\_schu67@uni-muenster.de}}
\thankstext{email2}{\texttt{dwenz@uni-muenster.de}}

\date{Received: date / Accepted: date}
% The correct dates will be entered by the editor

\maketitle
%\linenumbers

\begin{abstract}
This manuscript details the proof-of-concept of a small-scale cryogenic heat pump demonstrator, a technology designed to enable high-flow xenon distillation systems for the removal of \isotope[222]{Rn} in future liquid xenon observatories such as the XLZD experiment.
The heat pump demonstrator operates on a left-turning Clausius-Rankine cycle, utilizing xenon as a phase-chang\-ing working medium.
The design aims to fully hermetically separate the heat pump from the radon removal system, simplifying material cleanliness and maintenance compared to currently operating systems. 
Two demonstration tests were conducted at nominal pressures of \qty{3.3}{\bar} and \qty{4.3}{\bar}, utilizing a cold head and electrical heaters to mimic the behavior of a xenon distillation system.
In both measurements, the demonstrator achieved a cooling and heating power of {\CoolingPower} and {\HeatingPower}, respectively. This is sufficient to operate a small distillation system with a virtual purification mass flow of about {\VirtualPurificationFlow}, while consuming {\ElectricalPowerThreeBar} electrical power. 
Compered to currently operating applications using commercial cold heads driven by helium compressors, which typically require about \qty{6}{\kilo \watt} of electrical power, this is significantly lower.
The presented proof-of-concept heat pump demonstrator is further put into perspective with the currently planned XLZD experiment using a simplified scaling model.
This model indicates that a radon removal system with a purification mass flow of \qty{1600}{\kilo \gram \per \hour} and a required cooling and heating power of about \qty{60}{\kilo \watt} each, will be sufficient to cover a variety of different detector masses and background conditions.

\end{abstract}

\section{Introduction} 
\label{sec:introduction}
The search for dark matter in the form of weakly interacting massive particles (WIMPs) with masses above \qty{10}{GeV \per c^2} is dominated by experiments utilizing tonne-scale ($\mathcal{O}$(10 tonne)) xenon dual-phase time projection chambers (TPCs) \cite{xenon_wimp_sr1,lz_wimp,pandax_wimp}. 
The success of this detector technology lies in its excellent detection efficiency, scalability, ultra-low background, and the ability to discriminate between potential dark matter and background signals.
Thanks to the high density of liquid xenon (LXe), LXe TPCs enable the search for WIMPs in an almost background-free inner fiducial volume. Combined with additional veto detectors surrounding the TPC \cite{xenonnt_neutron_veto,lz_experiment}, and a stringent material screening \cite{xenonnt_screening}, material induced background signals from $\beta$-electrons and $\gamma$-rays are suppressed to a subdominant level.
Thus, only backgrounds which can enter the inner fiducial volume limit the dark matter sensitivity of current and future experiments such as XENONnT, LZ and PandaX, and XLZD, PandaX-xT, respectively \cite{xenonnt_projected_sensi,lz_experiment,PandaX_experiment,xlzd_design_book,PandaX_xT}.
There are two types of such backgrounds: unshieldable solar and atmospheric neutrinos, which leak into the WIMP signal region either through coherent elastic neutrino-nucleus scattering or neutrino-electron scattering \cite{xenonnt_cevns,pandax_cevns}, or radioactive noble elements, which diffuse into the inner fiducial volume and decay.
Although the former is an additional science signal in future liquid xenon observatories \cite{xlzd_design_book}, the latter must be mitigated.

Among all heavier noble elements (Ar, Kr, Xe, Rn), only \isotope[\mathrm{nat}]{Xe} contains exclusively ultra long-lived isotopes such as \isotope[124]{Xe} and \isotope[136]{Xe} with a half-life above $10^{21}\,$years, which are also utilized in the search for the neutrinoless double beta decay \cite{xenon_0nubb,XLZD_0vbb}.
However, through its extraction from air, and due to the emanation from detector materials, xenon contains the anthropogenic \isotope[85]{Kr} ($T_{1/2}=\qty{10.7}{years}$) and the short-lived \isotope[222]{Rn} ($T_{1/2}=\qty{3.8}{\day}$), respectively. 
The later originates from the decay chain of \isotope[238]{U}\footnote{The isotopes \isotope[219]{Rn} from \isotope[235]{U}, and \isotope[220]{Rn} from \isotope[232]{Th} only play a subdominant role thanks to their significantly shorter half-life \cite{xenonnt_rn_level_neutrino_floor}.}, commonly found in detector materials.
\isotope[85]{Kr} and the \isotope[222]{Rn} daughter \isotope[214]{Pb} produce low energetic beta decays which not only leak into the signal region for WIMPs, but also exhibit a similar spectral response as expected from electron scattering signals of solar-pp neutrinos \cite{xlzd_design_book}.

To mitigate these intrinsic backgrounds, LXe can be purified through cryogenic distillation as pioneered by XMASS in case of krypton removal \cite{XMASS}, and further developed and extended towards radon removal by the XENON collaboration \cite{xenonnt_rn_level_neutrino_floor,xenonnt_kr_removal,radon_removal_system}. 
The ra\-don removal system of the XENONnT experiment is composed of two components: a cryogenic distillation column, which exploits the differences in vapor pressure of two fluids to achieve separation through a repeated evaporation and condensation, and a radon-free xenon compressor that reduces external heating and cooling power requirements by operating the radon removal system in a heat-pump-like mode \cite{radon_removal_system}.
During the distillation, less volatile radon accumulates inside the LXe at the bottom of the distillation column, where it is trapped until it decays, while the purified ultra-clean gaseous xenon (GXe) is extracted from the top of the column.
The extracted clean gas is compressed by the radon-free compressors and condensed using the LXe inside the bottom part of the column.
This in-line heat-pump-like process reduces the external cooling power provided by liquid nitrogen (LN$_2$) from about \qty{3}{\kilo \watt} to about \qty{1}{\kilo \watt}, utilizing electrical power instead\footnote{The radon-free compressor is composed of 4 individual pumps with a power rating of \qty{1.5}{\kilo \watt} per drive. Three out of four compressors are typically running during operation \cite{xenonnt_rn_level_neutrino_floor,radon_removal_system}.}. 
At the same time the phase change of the condensing clean xenon provides about \qty{2}{\kilo \watt} of heating power to evaporate the LXe inside the bottom part of the column.
The remaining cooling and heating power of about \qty{1}{\kilo \watt} is provided through the evaporation of liquid nitrogen (LN$_2$) and electrical heater cartridges \cite{radon_removal_system}.
The reduction of \isotope[222]{Rn} in the detector strongly depends on the purification mass flow through the column, which sets the characteristic timescale for the radon removal relative to its decay within the active detector volume. 
With a continuous flow of \qty{62}{\kilo\gram\per\hour}, the XENONnT radon-removal system purifies the entire \qty{8.5}{\tonne} LXe volume of the experiment in \qty{5.7}{\day}.
This process enables a reduction in radon concentration by about a factor of four, to a baseline concentration of \qty{0.9}{\micro \becquerel \per \kilo \gram} \cite{xenonnt_rn_level_neutrino_floor}.

Although the existing concept achieves world-leading \isotope[222]{Rn} concentrations in XENONnT \cite{xenonnt_rn_level_neutrino_floor}, it is unlikely to meet the requirements of XLZD.
The custom-built radon-free compressors have a high requirement on material cleanliness, and moving parts cause abrasion on, e.g., the sealing of the pump's piston. 
This requires frequent maintenance and poses a constant risk of contaminating the ultra-pure xenon \cite{rn_compressor}.
In addition, the larger detector mass of up to \qty{104}{tonnes} and the more stringent radon background requirement of \qty{0.1}{\micro \becquerel \per \kilo \gram}, ten times lower than the expected neutrino background, necessitate a substantially larger purification flow and even tighter constraints on material purity.
Consequently, new technologies are required to realize a radon removal system for XLZD \cite{xlzd_design_book}.
Thus, this manuscript presents a proof-of-concept for a fully hermetically separated cryogenic heat-pump. 
The proposed design decouples the ultra-clean xenon circulation within the distillation column from an externally operated heat-pump loop, with heat being exchanged between the two systems through an oxygen-free copper heat exchanger \cite{heat_exchanger}. 
This approach preserves the advantage of reduced electrical power consumption by directly exploiting the latent heat stored within the processed ultra-clean LXe, rather than transferring the absorbed heat to the environment through cryocoolers or cold heads.  
To emulate the bottom and top sections of a distillation column, the proof-of-concept demonstrator is equipped with a cold head and resistive heaters, respectively.

The manuscript, is structured as follows: section \ref{sec:heatpump_demonstrator} explains the theoretical concept of the heat pump demonstrator, followed by its technical description, section \ref{sec:measurement} summarizes the performance measurement of the demonstrator under different initial conditions, and section \ref{sec:results} gives details about the resulting heating and cooling power provided by the system. Section \ref{sec:xlzd} then discusses the scaling to a XLZD-sized radon removal system and section \ref{sec:conclusion} concludes the manuscript.

\section{The heat pump demonstrator}
\label{sec:heatpump_demonstrator}
\subsection{Working principle and working medium}
\label{sec:working_principle}
\begin{figure*}[t]
\centering
\includegraphics[width=0.95\textwidth]{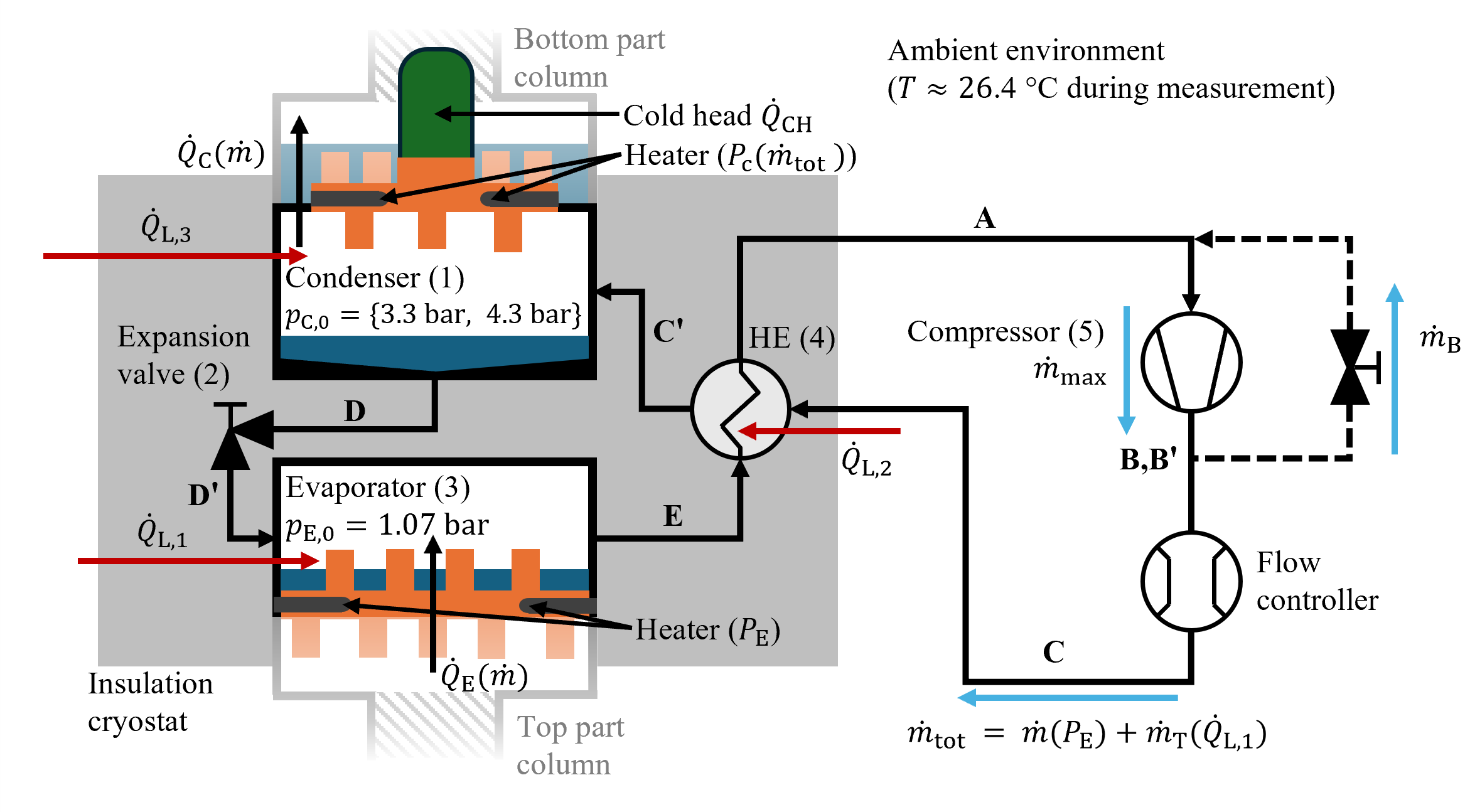}
\caption{Simplified schematic of the heat pump demonstrator representing the most important components: The condenser (1) provides the heating power $\dot{Q}_\mathrm{C}(\dot{m})$, which is absorbed by a ``cold reservoir'', defined by the fixed cooling power $\dot{Q}_\mathrm{CH}$ of a cold head (CH) depicted in green and an adjustable heating power $\dot{P}_\mathrm{C}(\dot{m}_\mathrm{tot})$ provided by counter-heating elements (dark gray). It is followed by an expansion valve (2) and the evaporator (3), in which additional heating elements serve as a ``heat reservoir'' providing an adjustable heating power $\dot{P}_\mathrm{E}$ equivalent to the cooling power $\dot{Q}_\mathrm{E}(\dot{m})$ of the heat pump. The gas then passes a gas-gas heat exchanger (HE) (4), and a compressor plus flow controller (5).
Components which are filled with LXe (indicated in blue) are mounted inside an insulation cryostat depicted as a gray region.
The remaining components are exposed to the ambient environment, which serves as a constant heat bath.
Red arrows indicate the external heat influx $\dot{Q}_j$, while black arrows show the heating and cooling power of the system.
Light blue arrows represent the mass flows $\dot{m}_j$.
Capital letters next to the lines indicate the different state points of the cycle as indicated in Fig. \ref{fig:rankine} and Table \ref{tab:state_points}.
When merged with a cryogenic distillation column, the cold head and heaters of the condenser are replaced by the bottom part of the column filled with LXe (light blue dashed).
The heaters of the evaporator are replaced by the top part of the column. 
The column parts are depicted as shaded components above the condenser and below the evaporator, respectively.}
\label{fig:heat_pump_sketch}
\end{figure*}

The thermodynamic design of the heat pump follows a left-turning Clausius-Rankine cycle with a phase-chang\-ing working medium. 
The cycle can be divided into five distinguished steps as illustrated in Fig. \ref{fig:heat_pump_sketch} and later discussed in Fig. \ref{fig:rankine}. 
Starting at (1), the working medium enters the condenser as a ``hot''\footnote{Note that ``hot'' and ``cold'' refer to the temperature relative to the operating temperature of the column around \qty{-93}{\celsius}. 
For the cryogenic heat pump presented here ``hot'' refers to a temperature of about \qty{-78}{\celsius}.} compressed gas, condenses and transfers heat to a ``cold  reservoir'', e.g., the bottom part of a distillation column or a cold head. 
The hot condensed liquid is then pushed through an expansion valve (2) which reduces the pressure through an isenthalpic expansion of the fluid, and thus cools it thanks to the Joule-Thomson effect. 
Subsequently, the ``cold'' liquid enters an evaporator (3) where it extracts heat from a connected heat reservoir through evaporation, e.g., the top part of a distillation column or electrical heaters. 
Afterwards, the evaporated cold gas passes through a heat exchanger (4) where it warms up to room temperature via an isobaric heat exchange before entering an external compressor (5). 
The gas exiting the compressor passes a flow controller before passing the heat exchanger and entering the condenser as ``hot'' compressed gas again, closing the cycle.

The specific design presented in this work was developed using a custom made calculation tool based on the thermodynamics libraries of CoolProp and TESpy \cite{Unkhoff2023,coolprop,tespy}. 
As working medium, a fluid with a liquid-gas transition around the operating temperature of the radon distillation column of about \qty{-93}{\celsius} must be chosen \cite{radon_removal_system}.
Further, the medium should only require small changes in operation pressure below \qty{5}{\bar} to achieve the required phase change, and should not pose any hazardous risk to meet the safety standards of underground laboratories, nor should the fluid show potential in global warming or ozone depletion.
Taking all requirements into consideration, only xenon itself was left as a suitable working medium. 
The working temperature of the condenser and the evaporator were designed to be \qty{-78}{\celsius} and \qty{-107}{\celsius}, respectively, corresponding to a nominal vapor pressure of $p_\mathrm{C,0}=\qty{4.3}{\bar}$ and $p_\mathrm{E,0}=\qty{1.07}{\bar}$.
This ensures a temperature gradient between the heat pump and the distillation column of about \qty{15}{\celsius} on each side and enables an efficient heat transfer between the two systems.
The ideal coefficient of performance $\mathrm{COP}_\mathrm{ideal}$ of a heat pump can be expressed as the ratio of the amount of extracted heat $Q_\mathrm{C}$ (or provided), over the work $W$ carried out by the compressor.
In such an ideal application, the COP for cooling ($\mathrm{COP}^\mathrm{c}$) and the COP for heating ($\mathrm{COP}^\mathrm{h}$) differ by one ($\mathrm{COP}^\mathrm{h}_\mathrm{ideal}=\mathrm{COP}^\mathrm{c}_\mathrm{ideal}+1$) \cite{TERBRAKE2002705}, and therefore the following discussion will be limited to $\mathrm{COP}^\mathrm{c}_\mathrm{ideal}$ only.
For a more application oriented approach, it is common to express the COP in terms of powers, allowing to expand the calculation into
\begin{equation}
\label{equ:cop_ideal}
    \mathrm{COP}^\mathrm{c}_\mathrm{ideal} = \frac{\dot{Q}_\mathrm{C}}{\dot{W}} = \frac{\dot{m}_\mathrm{max}\cdot \Delta h_\mathrm{e,d'}(p, Q)}{\dot{m}_\mathrm{max}\cdot \Delta h_\mathrm{b',a}(p, T)}\,,
\end{equation}
where $\dot{m}_\mathrm{max}$ is the maximum mass flow the compressor provides, and $\Delta h_\mathrm{i,j}$ the difference in the specific enthalpy between the state points i and j which are defined through the pressure p and the vapor quality Q or the temperature T at the given state point\footnote{The vapor quality indicates the fraction of vapor in a liquid-gas mixture. A quality of $Q=1$ means vapor only without any residual liquid.}. 
Please note that in equation \eqref{equ:cop_ideal} lower case letters indicate that the state points refer to the idealized design instead of measured state points, which are referenced by upper case letters.
The specific enthalpies of the state points a, b', d' and e are shown together with the measured values in Fig. \ref{fig:rankine}.
The $\mathrm{COP}^\mathrm{c}_\mathrm{ideal}$ of the designed heat pump was estimated to be {\COPDesignIdealFourBar}, assuming a gas temperature of $T_\mathrm{a}=T_\mathrm{A}=\qty{26.5}{\celsius}$ at the compressor inlet after passing the GXe-GXe heat exchanger, which corresponds to the laboratory temperature at the time of measurement\footnote{While originally designed for a condenser pressure of \qty{4.3}{\bar} a second measurement with a nominal condenser pressure of $p_\mathrm{C,0}=\qty{3.3}{\bar}$ was conducted as well. Here, the $\mathrm{COP}^\mathrm{c}_\mathrm{ideal}$ is slightly higher and expected to be {\COPDesignIdealThreeBar}.}.

This ideal COP can be linked to a measurable realistic $\mathrm{COP}^\mathrm{c}_\mathrm{real}$ through the product of different efficiencies $\eta_\mathrm{i}$
\begin{equation}
\label{eq:cop_real}
    \mathrm{COP}^\mathrm{c}_\mathrm{real} = \frac{\dot{m} \cdot \Delta h_\mathrm{E,D'}(p, Q)}{P_\mathrm{el}} = \prod_\mathrm{i} \eta_\mathrm{i} \cdot \mathrm{COP}^\mathrm{c}_\mathrm{ideal}\,,
\end{equation}
where $\dot{m}$ is the true mass flow through the heat pump, $P_\mathrm{el}$ is the electrical power consumed by the compressor, and $\Delta h_\mathrm{E,D'}(p, Q) \approx  \Delta h_\mathrm{e,d'}(p, Q)$ is the difference in specific enthalpy between the measured state points E and D'.
The maximum heating and cooling power of the proof-of-concept demonstrator was designed to drive a small krypton distillation column, which requires about \qty{50}{\watt} each, and is being developed in parallel. 

\subsection{Technical design}
\label{sec:technical_design}
\begin{figure*}[t]
\centering
\begin{subfigure}{0.49\textwidth}
\includegraphics[width=0.95\textwidth]{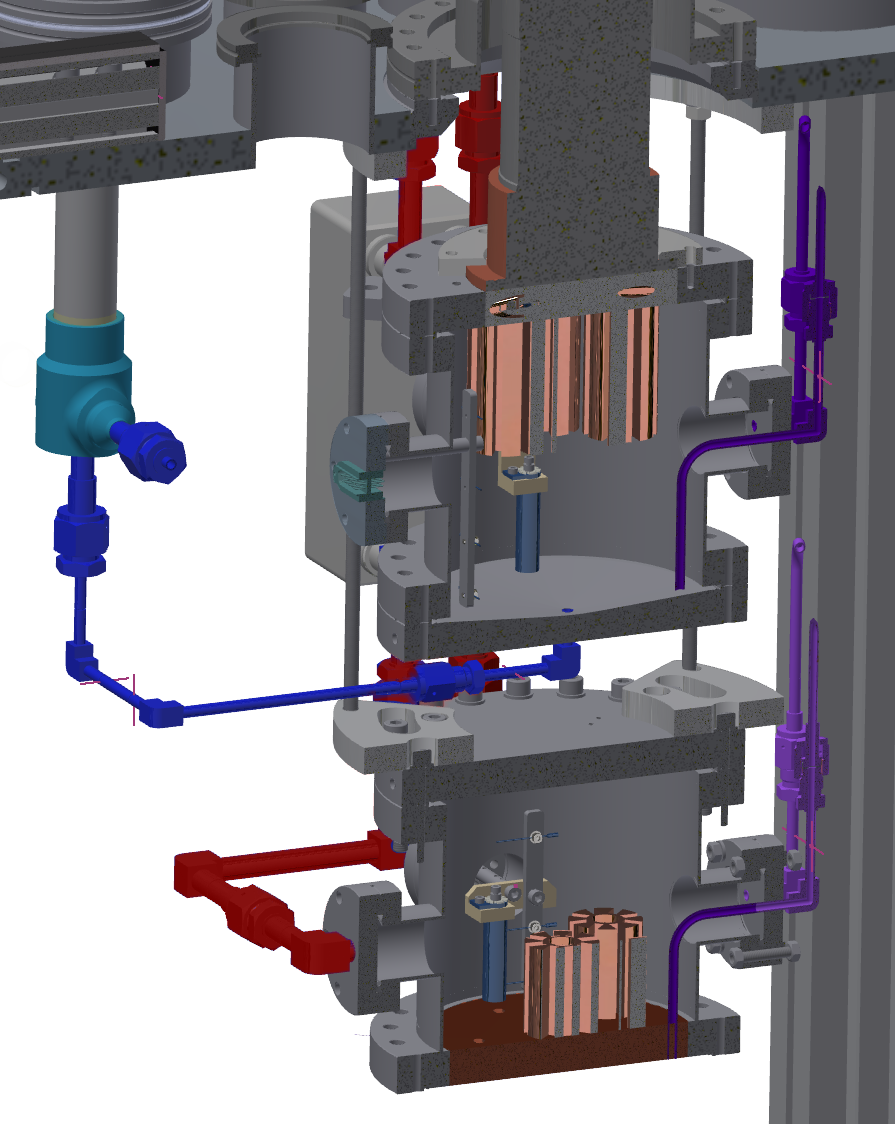}
\end{subfigure}
\begin{subfigure}{0.49\textwidth}
\includegraphics[width=0.95\textwidth]{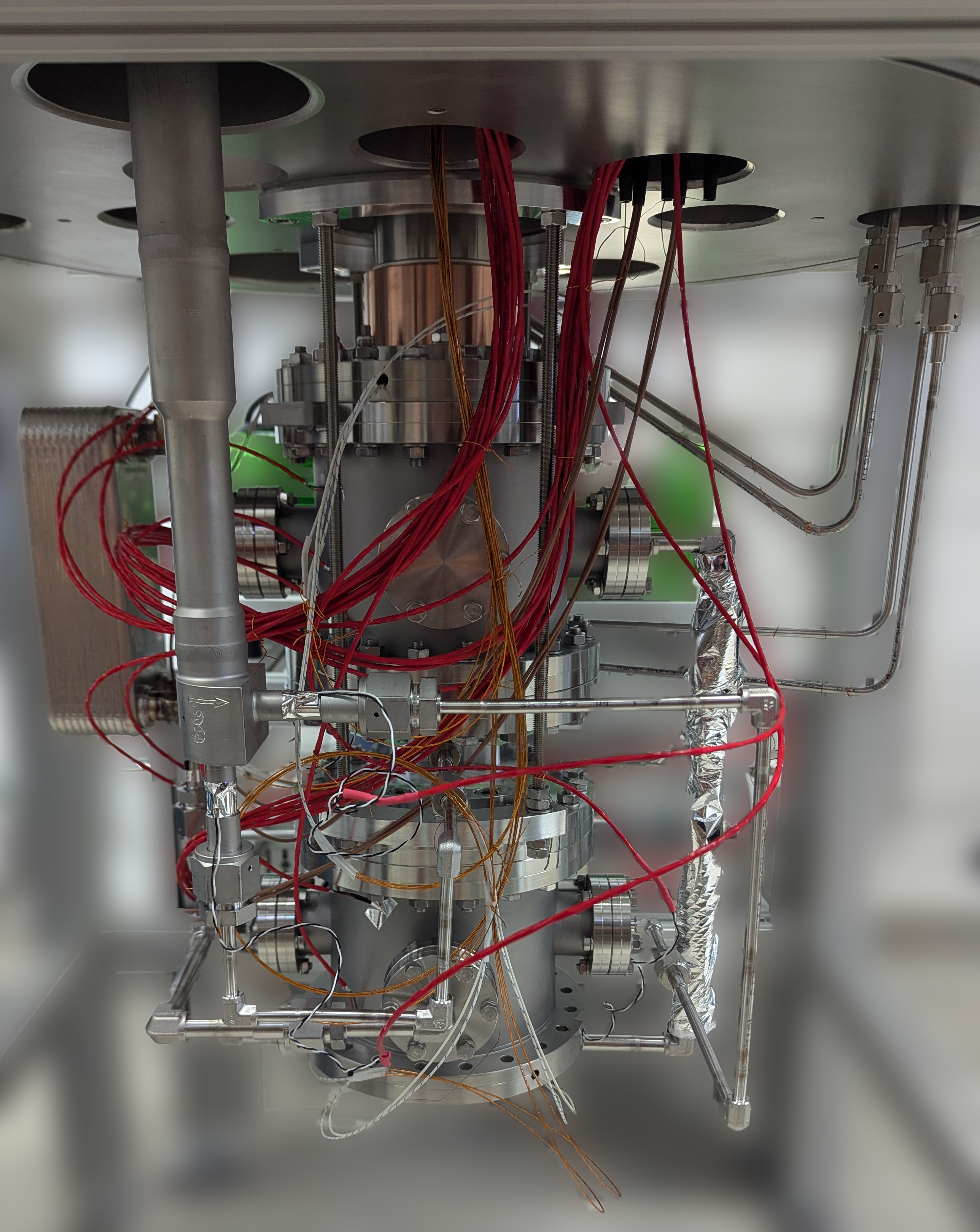}
\end{subfigure}
\caption{Left: CAD rendering of the heat pump demonstrator with a cutaway view into the condenser (top vessel) and evaporator (bottom vessel). The internal copper pins for improved heat transfer and the cylindrical capacitor level meters are visible. Pipes depicted in red and blue transport gaseous and liquid xenon, respectively. Pressure sensors are connected with condenser and evaporator through pipes colored in purple. The expansion valve body is highlighted in light blue. The outlet pipe of the expansion valve is connected to the bottom vessel at the cut-away CF40 connection. Right: Photograph of the cryogenic components installed inside the open insulation cryostat, showing the condenser, evaporator, and the connecting pipelines and cables. The image is rotated by about \qty{45}{\deg} with respect to the CAD model}
\label{fig:heat_pump_technical_design}
\end{figure*}

The technical design of the heat pump demonstrator is shown in Fig. \ref{fig:heat_pump_technical_design}.
Condenser and evaporator vessels have a CF160 standard diameter (\qty{150}{\milli \meter} inner diameter) with a height of \qty{160}{\milli \meter} and an inner height of \qty{140}{\milli \meter}, respectively. 
Each vessel features four CF40 feedthroughs at its sides to connect xenon supply lines and readout sensors.

The condenser vessel is closed at the top via a \qty{19.8}{\milli \meter} thick CF160 flange which thermally links the contained xenon gas with a cold head through an electro-welded copper disk with a diameter of \qty{115.2}{\milli \meter}.
The cold head (COOLPOWER 140i, DN 160 ISO-K) is driven by a helium compressor (COOLPAK 5000i), both are from the manufacturer Leybold and provide together a fixed cooling power of about \qty{250}{\watt} at \qty{-100}{\celsius} by consuming about \qty{6}{\kilo \watt} of electrical power \cite{LeyBoldManual}.
Together with three heater cartridges (S5105-1-4x21-2\-x125\-x300) of the manufacturer WEMA, which are placed inside holes at the side of the electro-welded copper disk, and a pro\-gram\-ma\-ble power supply from TDK Lambda (Z320-2), the cold head acts as a cold reservoir with an adjustable constant temperature. 
This mimics the bottom part of a radon distillation column.
The bottom of the condenser is closed with a CF160 stainless steel flange with a conical milling forming a funnel on the inside and a 1/4 inch liquid extraction port at its center. 
LXe leaving the condenser is pushed through a custom-made cryogenic expansion valve of the manufacturer SAMSON with a maximum flow factor of $k_\mathrm{vs}=\qty{0.004}{\cubic \meter \per \hour}$. 
The relaxed fluid enters the evaporator through one of the CF40 ports.
The evaporator is closed at the top with a CF160 blind flange and a \qty{19.8}{\milli \meter} thick copper disk at the bottom, which is directly welded into the CF160 vessel\footnote{The copper disk was directly electro-welded into the system to allow a later combination with the aforementioned krypton concentrator.}. 
Three additional heater cartridges, the same as for the condenser, are mounted into the electro-welded copper plate providing an adjustable heat load which mimics the top part of a radon distillation column as indicated in Fig. \ref{fig:heat_pump_sketch}.
Five copper pins following the design in \cite{heat_exchanger} are mounted to the copper plates of the condenser and evaporator to improve the heat transfer between xenon and the respective reservoir by increasing the surface area by \qty{264}{ \centi\square\meter} per pin. 
GXe exiting the evaporator warms to room temperature by cooling incoming GXe inside a commercial plate heat exchanger (EWT-BE4-13x20) from EWT Plattenw\"armetauscher, which provides a surface area of \qty{0.3}{\square\meter}.
The incoming GXe is supplied by a custom-made double membrane pump (N630.15.12) from KNF, which features PTFE-coated EPDM rubber membranes.
The pump consumes about \qty{350}{\watt} of electrical power, depending on its load, and provides a constant flow up to \qty{20}{slpm} (at $\Delta p=\qty{1}{\bar}$), depending on the pressure difference between inlet and outlet. 
The consumed electrical power was measured with a three-phase electrical power counter (Shelly Pro 3EM-3CT63).
A particle filter (High Purity Gas Filter DEF280FP11) from Mott is mounted downstream of the compressor, followed by a flow controller of the GM50A series of the manufacturer MKS\footnote{The reference temperature and pressure for the flow controller is specified by the manufacturer as \qty{0}{\celsius} and \qty{1013,25}{mbar}, while in this manuscript we use as reference \qty{20}{\degree} and \qty{1013,25}{mbar}. All flows were translated accordingly.}, which regulates the gas flow up to \qty{21.46}{slpm} depending on the load on the heat pump.
Given that the compressor provides a constant flow, residual gas is returned to the compressor inlet through a bypass line with an adjustable pressure relief valve (SS-RL4S8) from Swaglok. 
GXe and LXe are transported between components through electro-polished stainless steel pipes of the company Dockweiler with a diameter of 1/2 inch and 1/4 inch, respectively. 
All components are connected using face seal connections.

The thermal insulation of the cryogenic system is given through an insulation vacuum within a vessel measuring \qty{650}{\milli \meter} in diameter and \qty{577.5}{\milli \meter} in height.
The insulation vacuum is provided through a turbomolecular pump (TURBOVAC 350i) connected to a rough pump (TRIVAC D16 B), both from the company Leybold. 
The pressure of the insulation vacuum is monitored with a cold-cathode pressure gauge (VSM\-79DL) from Thyracont. 
During all operations, the insulation vacuum was typically in the order of \qty{e-7}{\milli \bar}.
An additional 10-layer insulation foil (COOLCAT 2NW) from the company Beyond Gravity is wrapped around all components containing LXe to further suppress external heat inflow. 

WIKA pressure sensors (WU20) are used to monitor the pressure of the evaporator and condenser, as well as the compressor inlet and outlet. 
Two silicon diodes (LS-DT-670D-CU) are mounted inside the electro-welded copper disk of the cold head and evaporator and are connected to a temperature-controller (LS-336), both of the company LakeShore.
In addition, PT-1000 resistance temperature detector (RTD) sensors (HEL-705-U-1-12-00) of the manufacturer Honeywell are mounted inside the condenser and evaporator to be directly submerged in GXe or LXe.
Additional PT-1000 RTDs (PP\-G102JA) from the company Littlefuse are mounted to the cryogenic piping inside the insulation vacuum.
The liquid level is monitored by a custom cylindrical capacitor level meter per vessel, with a height of \qty{50}{\milli \meter} and a change in capacity of \qty{5.88}{\pico \farad} per mm of liquid. 
Both level meters are read out via an UTI evaluation board of the company Smartec. 

\subsection{Slow control and monitoring}
\label{sec:slowcontrol}

The heat pump demonstrator is controlled by a central programmable logic controller (PLC) responsible for the sensor readout, auxiliary device information, logic driven control outputs and long term storage of parameter values. 
The Siemens PLC (S7-1200) is extended by one analog input (SM 1231), one analog output (SM 1232) and two RTD readout modules (SM 1231), featuring in total 8 analog input, 4 analog output and 16 RTD channels. 
Furthermore, additional readouts are realized via direct and indirect Ethernet/TCP IP connection to the LS-336 temperature-controller, powermeter, UTI evaluation board and the Z320-2 pro\-gram\-ma\-ble power supplies.
All PLC-controlled parameters are sent every second to a time series database using an InfluxDB platform.

The database runs on a Linux machine that also hosts a Grafana instance and a NodeRed server.
The Grafana service is primarily used for trend monitoring and accessing the history of each parameter.
NodeRed completes the slow control by directly communicating with the PLC and providing a self-developed supervisory control and data acquisition (SCADA) interface.
The interface shows the P\&ID of the heat pump demonstrator, including real-time updates of the current parameter values, and providing the possibility to control the state of all connected devices.
The NodeRed instance enables further remote controlled operations and an automated alarm system  sending notifications via e-mail and SMS.
An external server is constantly monitoring the availability of the alarm system and a dedicated uninterruptible power supply (UPS) for the heat pump demonstrator and its slow control devices takes over in case of a power cut.

Three separate PID control loops are configured via the NodeRed instance using self-defined control loops based on the Siemens PIDs. 
In total they mimic the behavior of a connected ``virtual'' distillation column and stabilize the liquid levels inside evaporator and condenser if the heat load of the virtual column changes.
The heat load of the virtual column is set directly by the electrical power $P_\mathrm{E}$ supplied to the evaporator heaters, thus defining the cooling power $\dot{Q}_\mathrm{E}(\dot{m})$ of the heat pump.
The first control loop manages the fixed temperature of the condenser. 
It works against the constant cooling power $\dot{Q}_\mathrm{CH}$ of the cold head by actively adjusting the power $P_\mathrm{C}(\dot{m}_\mathrm{tot})$ of the three counter-heaters mounted inside the electro-welded copper disk, maintaining a stable temperature gradient to the virtual distillation column.
Consequently, the heating power provided by the heat pump $\dot{Q}_\mathrm{C}(\dot{m})$ can be indirectly determined by measuring $P_\mathrm{C}(\dot{m}_\mathrm{tot})$.
The second control loop monitors the liquid level of the evaporator and adjusts the opening of the expansion valve if needed. 
It has a slow reaction time to reduce the stress on the expansion valve tip. 
The third loop is faster and adjusts the opening of the flow controller downstream of the compressor depending on the evaporator pressure. 
This indirectly links the heat load $P_\mathrm{E}$ of the virtual distillation column with the xenon mass flow $\dot{m}$ through the liquid level of the evaporator.

\section{Demonstration measurement}
\label{sec:measurement}
To characterize the behavior of the heat pump under different load conditions and to measure the provided heating and cooling power as a function of xenon flow, two measurement campaigns were carried out, defined by the nominal condenser pressure set to $p_\mathrm{C,0}=\qty{3.3}{\bar}$ and $p_\mathrm{C,0}=\qty{4.3}{\bar}$ at zero heat load ($P_\mathrm{E}=\qty{0}{\watt}$). 
The pressures were adjusted through the compressor bypass (dashed line in Fig. \ref{fig:heat_pump_sketch}) and the cold head temperature (\qty{-84.4}{\celsius} for a xenon vapor pressure of \qty{3.3}{\bar}, and \qty{-78.0}{\celsius} at \qty{4.3}{\bar}).
The inlet pressure of the compressor was controlled to be constant at \qty{1070\pm20}{\milli \bar} throughout the entire measurement.
The lower condenser pressure was chosen to test the performance of the expansion valve when accounting for the hydro-static pressure losses of about \qty{1}{\bar} caused in a \qty{3}{\meter} tall LXe column.
In both measurements, the electrical power $P_\mathrm{E}$ of the evaporator heaters was varied between \qty{0}{\watt} and \qty{130}{\watt}.

The two measurements showed no qualitative differences in the system's behavior, except for a systematic larger opening of the expansion valve during the measurement conducted at \qty{3.3}{\bar}. 
This is expected as the liquid flow through the expansion valve can be approximately described as
\begin{equation}
\label{eq:liquid}
\dot{m}_\mathrm{tot} = k_\mathrm{v} \cdot \sqrt{\frac{\rho \cdot \Delta p \cdot \qty{1000}{\kilo \gram \per \cubic \meter}}{\qty{1}{\bar}}},
\end{equation}
where $k_\mathrm{v}$ represents the flow factor of the valve for a given opening, $\Delta p$ the pressure drop across the valve in bar, and $\rho$ the fluid density in kg/m$^3$.
Across all set heat loads, the valve opening was consistently \qty{12\pm1}{\percent} larger than in the \qty{4.3}{\bar} measurement.
Since no qualitative difference was observed between the two measurements, only the \qty{3.3}{\bar} measurement will be discussed in the following, but the results for the \qty{4.3}{\bar} measurement  will be reported when applicable. 

\begin{figure*}[tb]
\centering
\includegraphics[width=0.99\textwidth]{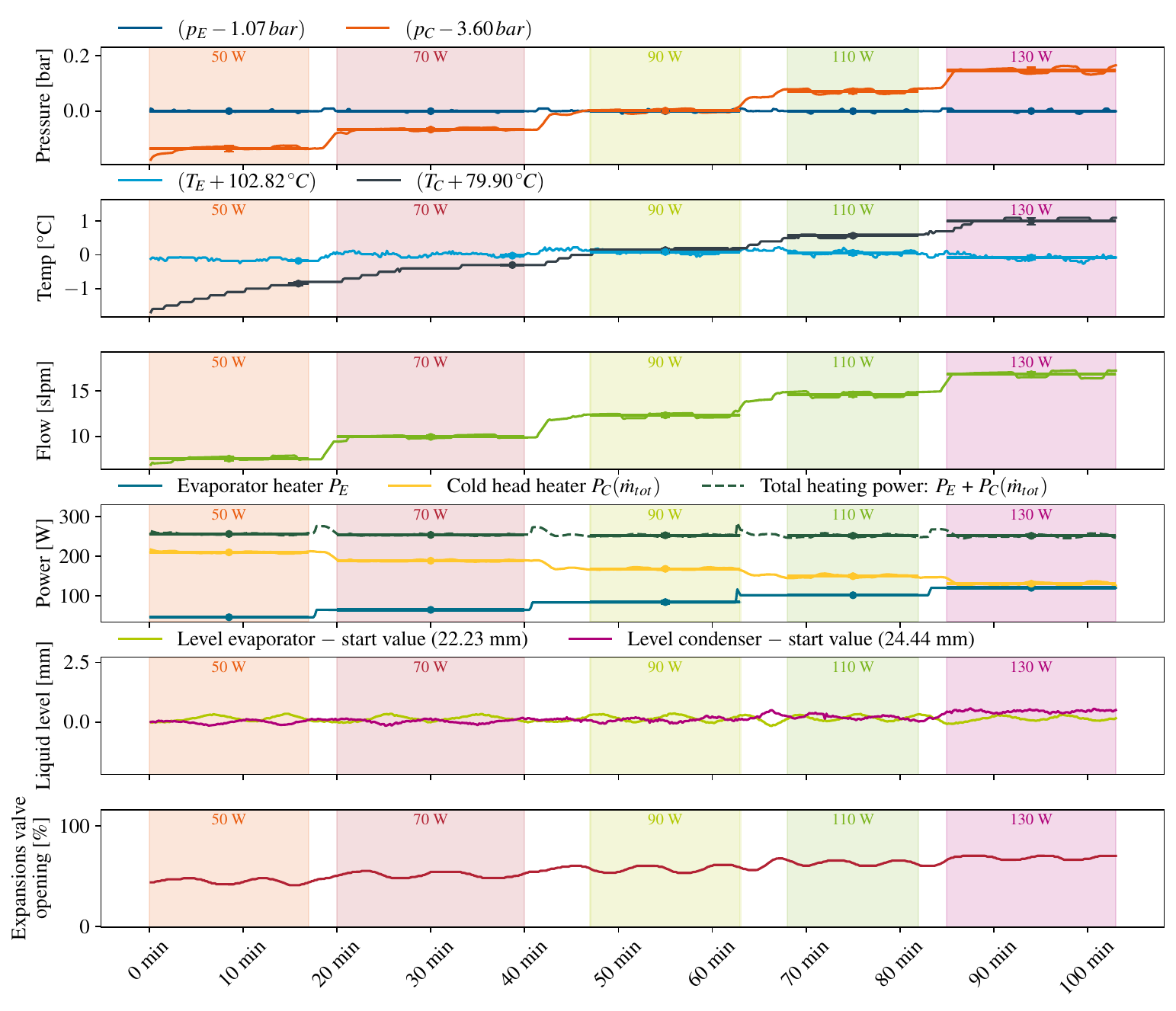}
\caption{System parameters as a function of time for the $p_\mathrm{C}(\qty{0}{\watt})=\qty{3.3}{\bar}$ performance measurement.
If applicable, either the first sensor reading or the average are subtracted from the shown parameters as indicated by the respective plot legend.
Colored shaded regions represent periods over which the different parameters were averaged for a given heat load $P_E$ ranging between \qty{90}{\watt} and \qty{130}{\watt}. 
The average for the \qty{50}{\watt} and \qty{70}{\watt} measurements were estimated as detailed in the text.
The average values are depicted as horizontal lines, and the standard deviation of each period is indicated by the error bar at the data point in the center of an averaging period}
\label{fig:slowcontrol_stability}
\end{figure*}

Fig. \ref{fig:slowcontrol_stability} illustrates the most important systems parameters for five different heat loads $P_\mathrm{E}$ between \qty{50}{\watt} and \qty{130}{\watt}.
To quantify the performance of the demonstrator, the heating $\dot{Q}_\mathrm{C}(\dot{m})$ and cooling power $\dot{Q}_\mathrm{E}(\dot{m})$ of the heat pump can be evaluated based on the mass flow $\dot{m}$ and the specific enthalpies determined by the measured pressures and temperatures at the different steady-state points of the Clausius-Rankine cycle.
These values can be then cross checked against measured electrical power of the condenser and evaporator heaters $P_\mathrm{E}$ and $P_\mathrm{C}(\dot{m}_\mathrm{tot})$.
The pressure $p_\mathrm{E}$ and temperature $T_\mathrm{E}$ inside the evaporator are nearly constant, while for the condenser, both temperature $T_\mathrm{C}$ and pressure $p_\mathrm{C}$ increase slightly with increasing heat load $P_\mathrm{E}$.
The heat load $P_\mathrm{E}$ and the power $P_\mathrm{C}(\dot{m})$, which counterbalances the constant cooling power of the cold head $\dot{Q}_\mathrm{CH}$, exhibit a perfectly anticorrelated behavior, with their sum remaining constant. 
This, indicates stable ``heat pumping'' from the evaporator to the condenser.

An unintentional feedback between the control loop of the expansion valve and flow controller led to an  oscillating opening of the expansion valve, and consequently small anti-correlated oscillations in the liquid levels (\qty{\pm0.1}{\milli \meter}), as well as small oscillations in the gas flow (\qty{\pm0.64}{slpm}), system pressures (\qty{\pm0.01}{\bar}) and temperature (\qty{\pm0.2}{\celsius}) which are all in phase. 
A faster control loop adjusted before the presented measurements allowed for a more stable operation, but required a constant actuation of the expansion valve motor.
To reduce the stress on the valve and its motor, the control was relaxed, accepting small residual oscillation as trade-off.
To minimize the impact of these oscillations on the relevant system parameters: temperature, pressure and flow, each measurement was averaged over a time period corresponding to at least two full oscillation periods of about \qty{15}{min}, after an initial waiting time of about \qty{5}{min} between heat load changes. 
Only the measured values for condenser pressure $p_\mathrm{C}$ and the condenser temperature $T_\mathrm{C}$ of the $P_\mathrm{E}=\qty{50}{\watt}$ and $P_\mathrm{E}=\qty{70}{\watt}$ measurements were estimated differently. 
At these measurements, the copper pins of the condenser have not reached an equilibrium temperature yet, given a larger load jump from \qty{0}{\watt} to \qty{50}{\watt}. 
Consequently, for these two measurements, $T_\mathrm{C}$ and $p_\mathrm{C}$ were estimated by averaging only over the last \qty{2}{min} of the measurement period. 
No significant impact on the derived heating and cooling power with respect to other neighboring load measurements was found in the later analysis (see also Fig. \ref{fig:cooling_heating}).
The statistical uncertainties of the averaged parameters are all negligible compared to their respective systematic uncertainties, and the differences between amplitudes and median of the oscillations are at the same order of magnitude as the uncertainties for the temperature and pressure readings. 
Only the flow exhibits a slightly larger difference which is accounted for as an additional systematic uncertainty in the analysis as discussed in section \ref{sec:results}. 

The systematic uncertainties of the sensor readings for the measured flow and pressures are provided by the manufacturer. 
The flow controller has a systematic uncertainty of \qty{\pm0.06}{slpm} in the range of \qty{0.43}{slpm} to \qty{4.29}{slpm}, and \qty{\pm0.21}{slpm} between \qty{4.29}{slpm} and \qty{21.46}{slpm}.
Pressure sensors are stated with a systematic uncertainty of \qty{\pm0.02}{\bar}.
The systematic uncertainty of the temperature measurement was evaluated with the system being in thermal equilibrium at room temperature, and comparing the groups of different PT-1000 sensors (inside the LXe vessel, insulation vacuum, or outside in contact with ambient air of the laboratory) against the more precise silicon diodes, which are readout using a 4-point measurement technique. 
This comparison revealed systematic offsets between silicon diodes and PT-1000 sensors. 
Acknowledging the intrinsic uncertainty of the reference diodes, measured by their spread in thermal equilibrium, a final conservative uncertainty was assigned. 
This results in an estimated systematic uncertainty of \qty{\pm0.2}{\celsius} for the silicon diodes and \qty{\pm0.5}{\celsius} for the PT-1000 sensors.
In addition, the measured electrical power of the heater cartridges is corrected by a constant factor of $\epsilon_\mathrm{R}=\qty{0.93}{}$ to account for resistive losses of the supply lines and solder joints, which were measured with a voltmeter.

Following the heat pump performance measurement, an additional test was conducted to determine the maximum compressor flow $\dot{m}_\mathrm{max}$ and the consumed electrical power $P_\mathrm{el}$ of the compressor, which are required to determine the COP of the system.
Both $P_\mathrm{el}$ and $\dot{m}_\mathrm{max}$, depend on the outlet pressure of the compressor and, therefore, on the heat load $P_\mathrm{E}$ and were measured under the same conditions as in previous measurements, with the inlet pressure of the compressor being kept constant at \qty{1070\pm20}{mbar}. 
At a pressure of \qty{3.75\pm0.02}{\bar}, which corresponds to the same pressure as the \qty{130}{\watt} measurement, a maximum flow of $\dot{m}_\mathrm{max}={\MaxCompressorFlowThreeBar}$ and $P_\mathrm{el}={\ElectricalPowerThreeBar}$  electrical power consumption were measured  ({\MaxCompressorFlowFourBar} and {\ElectricalPowerFourBar} at \qty{4.51\pm0.02}{\bar}).

\section{Demonstrator performance}
\label{sec:results}
To quantify the performance of the demonstrator, the heating $\dot{Q}_\mathrm{C}(\dot{m})$ and cooling power $\dot{Q}_\mathrm{E}(\dot{m})$ of the heat pump were evaluated under steady-state conditions determined through the pressures and temperatures of condenser, evaporator, and compressor inlet and outlet, and compared against the measured heating $\dot{Q}'_\mathrm{C}(\dot{m})$ and cooling power $\dot{Q}'_\mathrm{E}(\dot{m})$ determined by the electrical heater powers $P_\mathrm{E}$ and $P_\mathrm{C}(\dot{m}_\mathrm{tot})$ . 
The measured cooling power can be read off directly as $\dot{Q}'_\mathrm{E}(\dot{m})\approx\epsilon_\mathrm{R}\cdot P_\mathrm{E}$, while the measured heating power is determined indirectly through 
\begin{equation}
\label{equ:measured_cooling_power}
    \dot{Q}'_\mathrm{C}(\dot{m}) \approx \epsilon_\mathrm{R} \cdot (P_\mathrm{C}(\dot{m}_\mathrm{T}) - P_\mathrm{C}(\dot{m}_\mathrm{tot}))\,,
\end{equation}
where the total mass flow $\dot{m}_\mathrm{tot} = \dot{m}(P_\mathrm{E}) + \dot{m}_\mathrm{T}(Q_\mathrm{L,1})$ is the sum between the mass flow of the evaporated xenon $\dot{m}(P_\mathrm{E})$ for a given heat load $P_\mathrm{E}$, and $\dot{m}_\mathrm{T}(Q_\mathrm{L,1})$ the mass flow of the evaporated xenon due to thermal losses $Q_\mathrm{L,1}$ in the evaporator. 
The mass flow $\dot{m}_\mathrm{T}$ was determined to be {\MZeroFlowThreeBar} by measuring the flow at $P_\mathrm{E}=\qty{0}{\watt}$, corresponding to a thermal loss of {\HeatLossesThreeBar}.
For the $p_\mathrm{C,0}=\qty{4.3}{\bar}$ measurement the corresponding flow and thermal loss are {\MZeroFlowFourBar} and {\HeatLossesFourBar}, respectively.

Similar thermal losses are expected to also enter the system via the heat exchanger and the condenser ($Q_\mathrm{L,2}, Q_\mathrm{L,3}$ in Fig. \ref{fig:heat_pump_sketch}), but cancel out when determining the measured heating power via equation \eqref{equ:measured_cooling_power}, which can be expanded into 
\begin{align}
\label{equ:measured_cooling_power_expanded}
    \dot{Q}'_\mathrm{C}(\dot{m}) =& (\dot{Q}_\mathrm{CH} - \dot{Q}_\mathrm{L,2} - \dot{Q}_\mathrm{L,3} - \dot{Q}_\mathrm{C}(\dot{m}_\mathrm{T})) \\
    &- (\dot{Q}_\mathrm{CH} - \dot{Q}_\mathrm{L,2} - \dot{Q}_\mathrm{L,3} - \dot{Q}_\mathrm{C}(\dot{m}) - \dot{Q}_\mathrm{C}(\dot{m}_\mathrm{T}))\,, \nonumber
\end{align}
as long as $Q_\mathrm{L,2}$ and $Q_\mathrm{L,3}$ are not flow dependent. 
The validity of this assumption is confirmed by the fact that the total power $P_C(\dot{m}_\mathrm{tot}) + P_E(\dot{m})$ in Fig. \ref{fig:slowcontrol_stability} remains constant under different heat loads $P_E(\dot{m})$.

\begin{figure*}[t]
    \begin{subfigure}{0.5\textwidth}
        \includegraphics[width=1\textwidth]{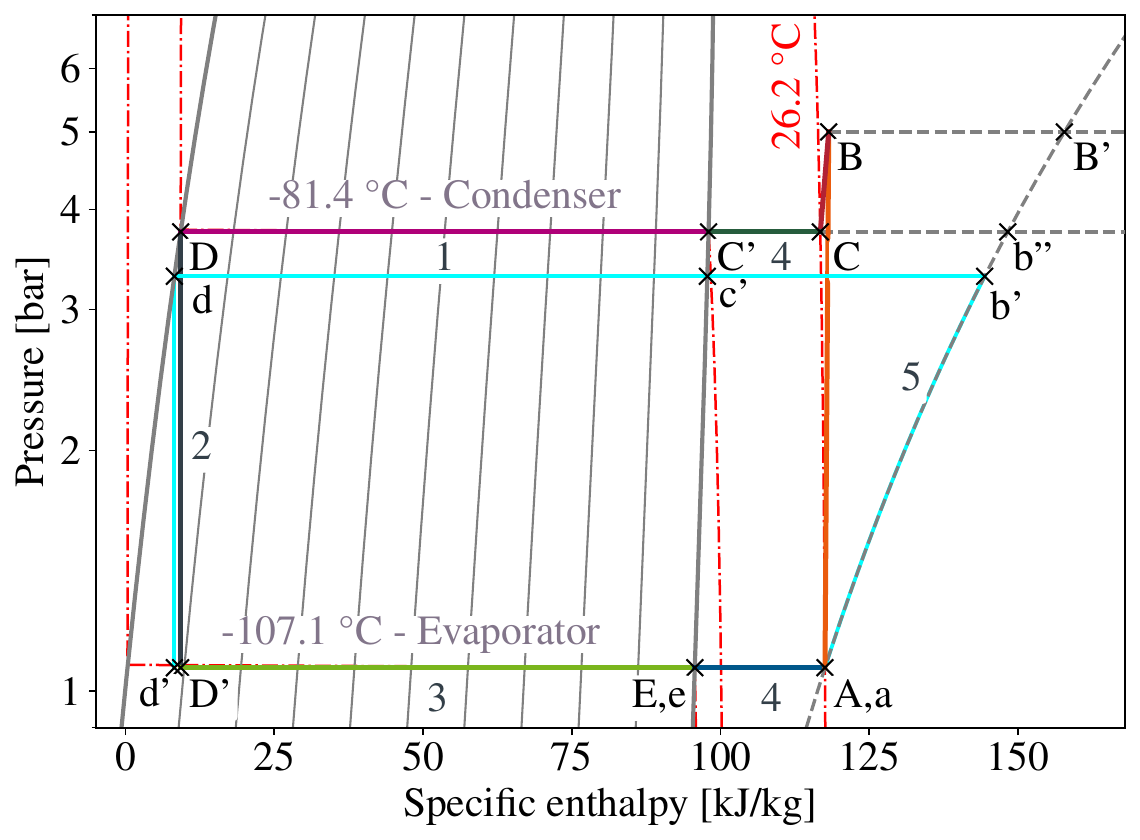}
        \end{subfigure}
    \begin{subfigure}{0.5\textwidth}
        \includegraphics[width=1\textwidth]{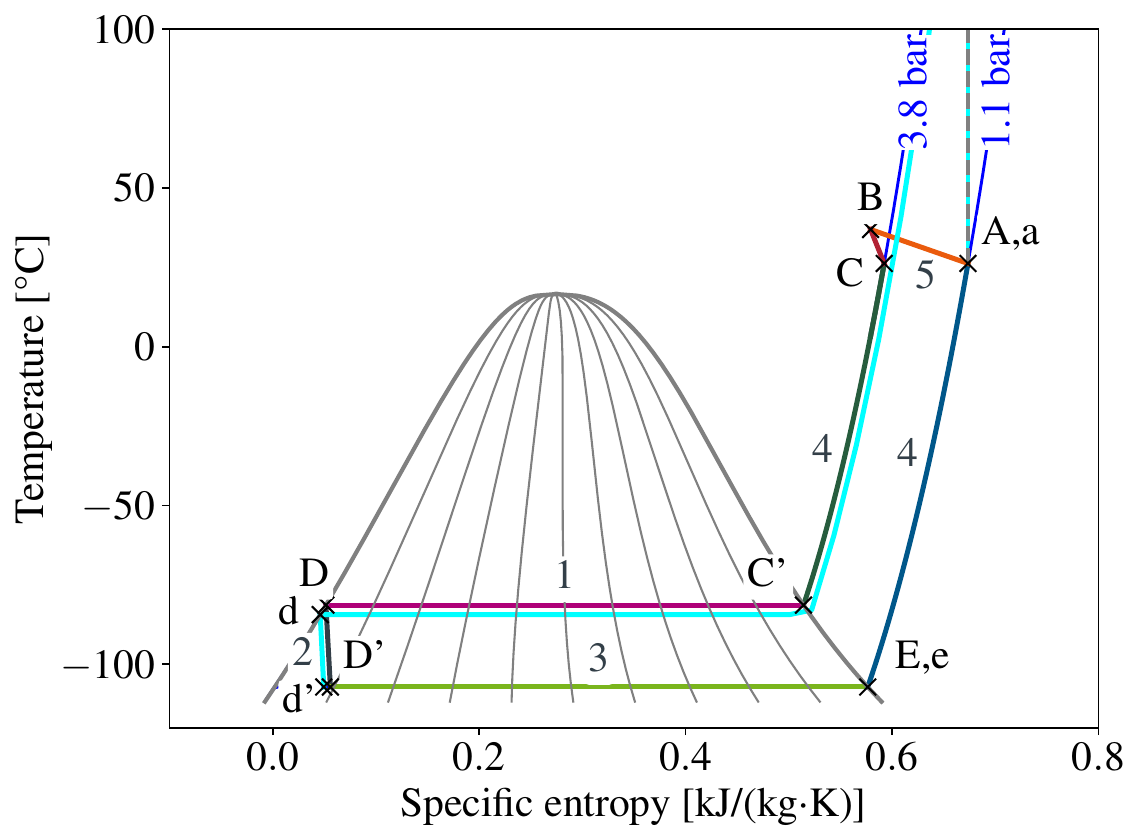}
    \end{subfigure}
    \caption{Pressure-enthalpy (left) and temperature-entropy (right) diagram of the Clausius-Rankine cycle for the \qty{130}{\watt} measurement. The two figures illustrate the different states of the heat pump's thermodynamic cycle. Measured state points are referenced as upper case letters, while design values use lower case letters. The gray dashed lines indicate the assumed ideal compression described in the text. Cyan colored lines show the idealized design cycle as reference. Red dashed-dotted lines show isothermal lines of different temperatures. Gray lines indicate the steam quality in \qty{10}{\percent} steps. The numbers correspond to the process steps depicted in Fig. \ref{fig:heat_pump_sketch}. Two isobaric lines are depicted in the right hand side figure in blue. The state points B', b' and b'' are omitted in the figure on the right hand side}
    \label{fig:rankine}       
\end{figure*}

Fig. \ref{fig:rankine} shows the different states (A to E) of the Clausius-Rankine cycle in a pressure-enthalpy and tem\-per\-a\-ture-entropy diagram for the \qty{130}{\watt} measurement.  
The specific enthalpy and specific entropy of the state points were determined with CoolProp \cite{coolprop}, and additional state points not accessible through direct measurements, denoted in the following discussion as primes, were determined.
All state points are also summarized in Table \ref{tab:state_points}.
\begin{table*}[tb]
\centering
\caption{Average thermodynamic state points, averaged over the $\approx\SI{15}{min}$ period of two pressure oscillations, for the \qty{130}{W} measurement. Measured values are indicated in bold font. The specific enthalpy, specific entropy and fluid phase are calculated from the measured pressure and temperature using the CoolProp library as explained in the text}
\label{tab:state_points}
\begin{tabular}{c c c c c c}
\toprule
\textbf{State Point} & 
\shortstack{\textbf{Pressure p} \\ \textbf{[bar]}} & 
\shortstack{\textbf{Temperature T} \\ \textbf{[\textdegree{}C]}} & 
\shortstack{\textbf{Specific enthalpy h} \\ \textbf{[kJ/kg]}} &
\shortstack{\textbf{Specific entropy s} \\ \textbf{[J/(kg K)]}} &
\shortstack{\textbf{Vapor quality Q} \\ ~} \\
\midrule
A  & \textbf{1.07$\pm$0.02} & \textbf{26.4$\pm$0.5} & $117.6\pm0.1$ & 674$\pm$1 & 1 \\
B  & \textbf{5.00$\pm$0.02} & \textbf{36.9$\pm$0.5} & $118.1\pm0.1$ & 579$\pm$1 & 1 \\
B'  & \textbf{5.00$\pm$0.02} & $282\pm5$ & $157.9\pm0.8$ & 674$\pm$2  & 1 \\
C  & \textbf{3.75$\pm$0.02} & \textbf{26.4$\pm$0.5} & $116.8\pm0.1$ & 592$\pm$1 & 1 \\
C'  & \textbf{3.75$\pm$0.02} & $-81.4\pm0.2$ & $98.0\pm0.01$ & 514.2$\pm$0.2 & 1 \\
D  & \textbf{3.75$\pm$0.02} & $-81.4\pm0.2$ & $9.27\pm0.05$ & 51.5$\pm$0.3 & 0 \\
D'  & \textbf{1.07$\pm$0.02} & $-107.1\pm0.2$ & $9.27\pm0.05$ & 54$\pm$6 & $0.09\pm0.01$ \\
E  & \textbf{1.07$\pm$0.02} & $-107.1\pm0.2$ & $95.69\pm0.04$ & 576$\pm$1 & 1 \\
\bottomrule
\end{tabular}
\end{table*}
To indicate the full cycle, a \qty{100}{\percent} efficient adiabatic compression between point A and B' is shown, using the temperature and pressure measured at point A and the pressure measured at point B as input.
The pressure drop between B and C following the compression is caused by the flow controller and a particle filter.
The state points C' and D are determined by the saturated pressure of the condenser and the vapor quality of Q=1 and Q=0, respectively. 
To estimate the vapor quality of the xenon after its expansion (D to D'), the expansion was assumed to be isenthalpic. 
To verify this assumption, the cooling power calculated for an ideal expansion $\dot{Q}_\mathrm{E}(\dot{m})$ was compared with the measured cooling power $\dot{Q}'_\mathrm{E}(\dot{m})$. 
This comparison yielded a ratio of $ 0.99\pm0.04$, consistent with an ideal expansion.
Based on this, point D' was determined as the intersection of the isenthalpic line from point D and the isobaric line from point E which is determined by the saturated vapor pressure of the evaporator and a vapor quality of Q=1.

\begin{figure*}[t]
    \begin{subfigure}{0.49\textwidth}
        \includegraphics[width=0.95\textwidth]{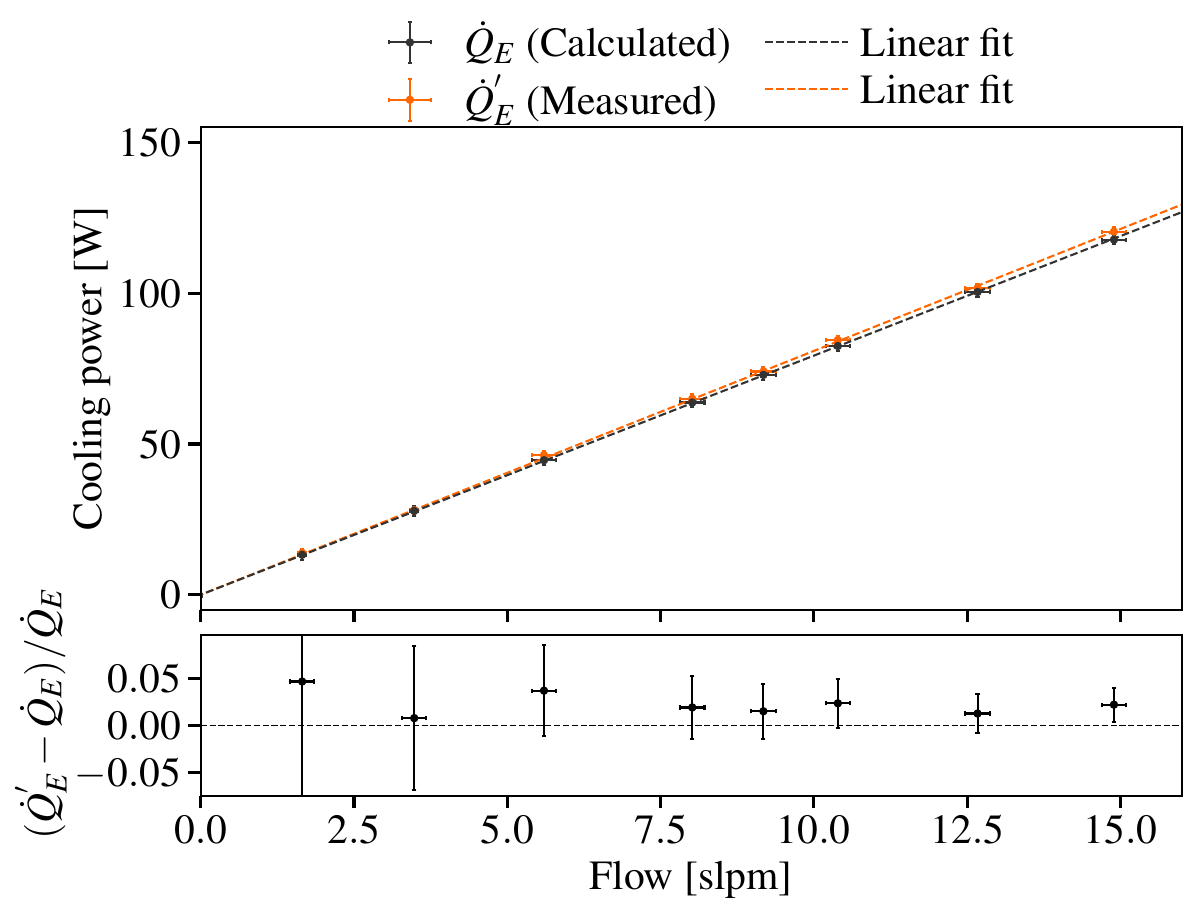}
        \end{subfigure}
    \begin{subfigure}{0.49\textwidth}
        \includegraphics[width=0.95\textwidth]{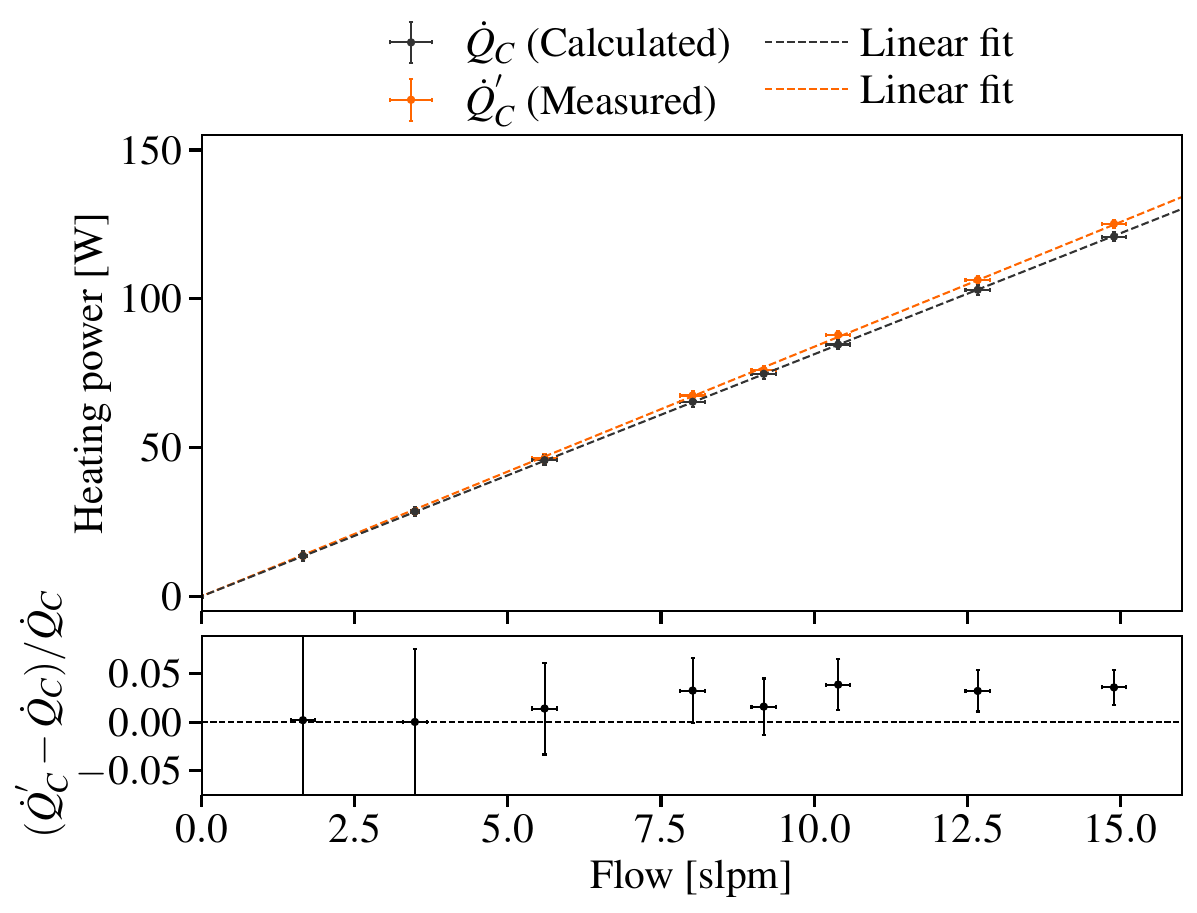}
    \end{subfigure}
    \caption{The figures on the left and right-hand side show the measured and calculated cooling and heating powers as a function of the corrected xenon mass flow for the \qty{3.3}{bar} measurement, respectively. The lower panel shows the relative difference between measured and calculated heating and cooling power.}
    \label{fig:cooling_heating}       
\end{figure*}

The state points E,D' and C',D are then used to estimate the cooling $\dot{Q}_\mathrm{E}(\dot{m})$ and heating $\dot{Q}_\mathrm{C}(\dot{m})$ power by multiplying the difference in specific enthalpy $(\Delta h_\mathrm{E,D'}$, $\Delta h_\mathrm{C',D})$, with the mass flow $\dot{m}$.
The resulting cooling and heating powers for the \qty{3.3}{\bar} measurement are compared against their measured counterparts in Fig. \ref{fig:cooling_heating}. 
Both measured and estimated cooling and heating power exhibit a linear dependence on the mass flow $\dot{m}$, consistent with the expected thermodynamic behavior of a steady-state Clausius-Rankine cycle.
The measured cooling and heating power show a systematic \qty{2}{\percent} bias with respect to their calculated counterparts. 
This might indicate either constant residual thermal losses, which have not been corrected for, or nonlinear effects caused by the oscillating system parameters. 
To acknowledge a potentially increased systematic uncertainty caused by the oscillations of the control loop, we added the half-difference between the estimated median and the oscillation amplitudes for the mass flow reading as an additional systematic uncertainty.
The difference between median and amplitude for the other sensor readings have a negligible impact on the estimated cooling and heating powers.
The resulting maximum cooling and heating power for the \qty{3.3}{\bar} measurement are $\dot{Q}_\mathrm{E}$={\CoolingPower} and $\dot{Q}_\mathrm{C}$={\HeatingPower} at a flow rate of $\dot{m}$={\MFullWattFlowThreeBar}, respectively. The same powers were observed at the \qty{4.3}{\bar} measurement, but at a flow rate of {\MFullWattFlowFourBar}.
Using the computed cooling power and the measured power of the compressor, the resulting realistic coefficient of performance is calculated using equation \eqref{eq:cop_real} and is found to be {\COPMeasuredRealThreeBar} for both measurements, which is significantly lower than the expected $\mathrm{COP}^\mathrm{c}_\mathrm{ideal} =${\COPDesignIdealThreeBar} for the $p_\mathrm{C,0}=$\qty{3.3}{\bar} measurement ($\mathrm{COP}^\mathrm{c}_\mathrm{ideal} =${\COPDesignIdealFourBar} for $p_\mathrm{C,0}=$\qty{4.3}{\bar}).

To identify the main cause for the efficiency loss we divided our loss in five terms
\begin{equation}
\label{eq:cop_real_expanded}
    \mathrm{COP}^\mathrm{c}_\mathrm{real} = \eta_\mathrm{T} \cdot \eta_\mathrm{B} \cdot \eta_\mathrm{C} \cdot \eta_\mathrm{P} \cdot \eta_\mathrm{P,HE} \cdot \mathrm{COP}^\mathrm{c}_\mathrm{ideal}\,,
\end{equation}
where $\eta_\mathrm{T}$ represents the efficiency loss due to thermal losses, $\eta_\mathrm{B}$ losses due to the compressor bypass, $\eta_\mathrm{C}$ losses due to the compressor choice, and $\eta_\mathrm{P}$, $\eta_\mathrm{P,HE}$ losses due to pressure losses with respect to the ideal design.
The specific enthalpies $h_\mathrm{i}$ are shown for the respective state points $i$ in Fig. \ref{fig:rankine}.
Starting from the right hand side the first loss term $\eta_\mathrm{P,HE}$ can be expressed as 
\begin{equation}
\label{eq:eta_p_he}
    \eta_\mathrm{P,HE} = \frac{\dot{m}_\mathrm{max} \cdot \Delta h_\mathrm{b',a}}{\dot{m}_\mathrm{max} \cdot \Delta h_\mathrm{b'',a}}= {\EtaHeatExchanger}\, 
\end{equation}
where $\Delta h_\mathrm{i,j}$ represents the difference in specific enthalpy between the state points $i$ and $j$.
It accounts for the higher pressure inside the condenser $p_\mathrm{C}$ at the state-point C' compared to the ideal design c', and the associated higher compressor work. 
This higher pressure correlates with the mass flow $\dot{m}$, as it can be seen in Fig. \ref{fig:slowcontrol_stability}, while within each load setting the pressure $p_\mathrm{C}$ remained constant.
This indicates an insufficient thermal conductivity between the condenser and the cold head, which requires a higher temperature gradient and thus a higher saturation pressure at higher mass flows. 
Based on previous works in \cite{heat_exchanger} we know that highly efficient LXe-bath type heat exchangers can be built and believe that this issue is mostly limited to the presented prototype. 
Thus, we decided to separate off this inefficiency term from the more general $\eta_\mathrm{P}$ term  
\begin{equation}
\label{eq:eta_p}
    \eta_\mathrm{P} = \frac{\dot{m}_\mathrm{max} \cdot \Delta h_\mathrm{b'',a}}{\dot{m}_\mathrm{max} \cdot \Delta h_\mathrm{B',A}}= {\EtaPressure}\,,
\end{equation}
which accounts for the pressure losses between state-points B and C resulting from the flow controller and particle filter compared to the ideal design. 
This pressure loss will be more difficult to mitigate and might also remain in a final XLZD sized design. 
The third loss term characterizes the inefficiency of the compressor by comparing the measured compression power with the consumed electrical power
\begin{equation}
\label{eq:eta_c}
    \eta_\mathrm{C} = \frac{\dot{m}_\mathrm{max} \cdot \Delta h_\mathrm{B',A}}{P_\mathrm{el}}= {\EtaCompressor}\,.
\end{equation}
This efficiency loss is found to be the highest and can be reduced by the choice of a more problem-specific gas compressor. 
The efficiency loss $\eta_\mathrm{B}$ accounts for the diverted bypass flow $\dot{m}_\mathrm{B}$ reducing the maximal available mass flow $\dot{m}_\mathrm{max}$ to ensure a stable operation. 
It is given by
\begin{equation}
\label{eq:eta_b}
    \eta_\mathrm{B} = \frac{\dot{m}_\mathrm{max} - \dot{m}_\mathrm{B}}{\dot{m}_\mathrm{max}}= {\EtaBypass}
\end{equation}
and strictly speaking only valid for the measurement at \qty{130}{\watt}. 
At smaller heat loads, this inefficiency increases as more gas is circulated through the bypass. 
For an XLZD sized design, this fact plays less of a role as the system will be designed for one specific high mass flow. 
The last inefficiency is induced by thermal losses which can be partially represented by thermal mass flow $\dot{m}_\mathrm{T}$, leading to an effective inefficiency 
\begin{align}
\label{eq:eta_b}
    \eta_\mathrm{T} = \frac{\dot{m}}{\dot{m}_\mathrm{max} - \dot{m}_\mathrm{B}} &= \frac{\dot{m}_\mathrm{max} - \dot{m}_\mathrm{B} - \dot{m}_\mathrm{T}}{\dot{m}_\mathrm{max} - \dot{m}_\mathrm{B}} \nonumber \\ &= {\EtaThermal}\,.
\end{align}
Not included in this inefficiency are the heat losses $\dot{Q}_\mathrm{L,2}$ and $\dot{Q}_\mathrm{L,3}$ (see Fig. \ref{fig:heat_pump_technical_design}),  which are compensated for by the constant cooling power $\dot{Q}_\mathrm{CH}$ of the cold head, and cannot be measured directly by the presented prototype.
Based on already operating distillation systems, we know that heat losses do not scale with the purification flow\cite{xenonnt_kr_removal,radon_removal_system}, and thus we expect this inefficiency to decrease for a larger system.
The total efficiency of the demonstrator $\eta_\mathrm{Carnot}$ defined as the measured $\mathrm{COP}^\mathrm{c}$ relative to the $\mathrm{COP}^\mathrm{c}$ of the Carnot limit is about 5\%, which is at the lower end for typical cryocoolers which range between \qty{10}{\percent} and \qty{20}{\percent} \cite{TERBRAKE2002705}. 

While the performance of a heat pump is typically characterized either by its heating $\dot{Q}_\mathrm{C}(\dot{m})$ or its cooling power $\dot{Q}_\mathrm{E}(\dot{m})$, the application of cryogenic distillation requires both simultaneously for the condensation and evaporation of the xenon inside the column. Therefore, we introduce a more application-oriented performance measure, which is defined as the ratio of the total usable thermal power to the consumed electrical power:
\begin{equation}
\mathrm{COP}_\mathrm{dist}^\mathrm{hc} = \frac{\dot{Q}_\mathrm{C}(\dot{m}) + \dot{Q}_\mathrm{E}(\dot{m})}{P_\mathrm{el}}.
\end{equation}
For the presented system, this performance factor is found to be {\COPMeasuredDistThreeBar} for both measurements.

\section{Projections for the XLZD experiment}
\label{sec:xlzd}

To place the performance of the presented heat pump demonstrator into the context of the planned XLZD experiment, an order-of-magnitude estimate was made for the required liquid flow and the associated cooling and heating power of a future radon removal system. 
These estimates are based on simplified scaling assumptions regarding the radon emanation rates and detector mass scaling, as well as simplified enthalpy-flow scaling for the cooling and heating power requirements.
In XENONnT the radon concentration before active removal was measured to be \qty{3.62\pm0.18}{\micro \becquerel \per \kilo \gram} \cite{radon_removal_system}. 
Scaling this value according to the volume-to-surface ratio which varies as $m_\mathrm{T}^{-1/3}$ ($m_T$ is the to be purified detector mass), yields an expected radon concentration of about \qty{1.8}{\micro \becquerel \per \kilo \gram} for the nominal detector mass of $m^\mathrm{XLZD}_\mathrm{nominal}=\qty{78}{tonne}$\footnote{In the opportunity design, which aims for a detector mass of $m^\mathrm{XLZD}_\mathrm{opportunity}=\qty{104}{tonne}$, the reduction increases slightly, but this improvement is neglected in the following estimates.}\cite{xlzd_design_book,xenonnt_rn_level_neutrino_floor}.
Assuming further that XLZD will achieve an additional reduction in radon concentration by a factor of three through passive and active mitigation strategies like more stringent material selection, surface treatment, cleanliness, and radon tagging \cite{xlzd_design_book}, the remaining radon concentration of \qty{0.6}{\micro \becquerel \per \kilo \gram} is still a factor of six larger than XLZD's final goal of \qty{0.1}{\micro \becquerel \per \kilo \gram}. 

Based on the radon removal model discussed in \cite{xenonnt_rn_level_neutrino_floor,radon_removal_system}, radon sources can be divided into three sub-types called type 1a, 1b and 2 depending on their location with respect to the radon removal system. 
Type 1a sources emanate radon directly into the LXe and thus must be removed from the liquid. 
Type 1b sources emanate into GXe and can be efficiently removed through an enforced gas extraction flow away from the liquid xenon volume. 
Type 2 sources are upstream of the radon removal system and thus do not play any role in the sizing of a future distillation system as they are removed with a \qty{100}{\percent} efficiency. 
In XENONnT, the radon source were almost evenly split between type 1a and type 1b sources. 

The reduction of type 1a sources is more challenging than the removal of type 1b sources. 
In XENONnT, the extraction efficiency of type 1b sources was measured to be about \qty{90}{\percent}, reducing the overall radon content in XENONnT by a factor of 2 for a moderate gas flow of \qty{20}{slpm} \cite{xenonnt_rn_level_neutrino_floor}. 
For the reduction of type 1a sources much larger liquid flows are required, as the time constant $\lambda_\mathrm{dist}$, in which the entire LXe volume of the experiment must be purified, is determined by 
\begin{equation}
    r = \frac{\lambda_\mathrm{Rn} + \lambda_\mathrm{dist.}}{\lambda_\mathrm{Rn}}
\end{equation}
where $r$ is the desired reduction factor, and $\lambda_\mathrm{Rn}=0.18\,\mathrm{d}^{-1}$ the \isotope[222]{Rn} decay constant \cite{radon_removal_system}.
In XENONnT, a flow of \qty{200}{slpm} was required to provide a radon reduction by another factor of $r=2$ \cite{xenonnt_rn_level_neutrino_floor}.
Since it is unclear if the radon sources will again be evenly split between type 1a and type 1b source, two different scenarios are considered in the following. 
In the conservative scenario, we assume that only $1/3$ of the \qty{0.6}{\micro \becquerel\per\kilo\gram} is emanated by type 1b source, while the nominal scenario assumes a similar even distribution as in XENONnT.
This implies that in the nominal (conservative) scenario a reduction factor of 2 (1.5) through gaseous extraction can be achieved, which requires the liquid extraction to provide the remaining reduction factor of $r=3$ ($r=4$) to achieve a total radon reduction by a factor of 6. 
These reduction factors can be achieved if the xenon volume of the detector $m^\mathrm{XLZD}_\mathrm{i}$ is exchanged at an exchange rate of $\lambda_\mathrm{dist}=0.36\,\mathrm{d}^{-1}$ ($\lambda_\mathrm{dist}=0.54\,\mathrm{d}^{-1}$). 
The resulting required liquid flows, as well as the required heating and cooling powers for such a system, are summarized for the nominal and opportunity detector design of the XLZD experiment in Table \ref{tab:liquid_flows} \cite{xlzd_design_book}.
\begin{table*}[tb]
    \centering
    \caption{Required liquid purification flow without and with column reflux, as well as resulting required cooling and heating power for the nominal and conservative radon emanation scenarios and the nominal and opportunity XLZD detector masses $m^\mathrm{XLZD}_\mathrm{i}$. The power values refer only to the power required to either evaporate or liquefy the xenon inside the column (including the reflux). The total required power is thus twice as large}
    \label{tab:liquid_flows}
    \begin{tabular}{c | c c c c c c}
        \toprule
        \textbf{Scenario} & \multicolumn{3}{c}{\textbf{$m^\mathrm{XLZD}_\mathrm{nominal}=\qty{78}{tonne}$}} & \multicolumn{3}{c}{\textbf{$m^\mathrm{XLZD}_\mathrm{opportunity}=\qty{104}{tonne}$}} \\
        \toprule
         &
        \shortstack{\textbf{Purification} \\ \textbf{[kg/h]}} &
        \shortstack{\textbf{with reflux} \\ \textbf{[kg/h]}} &
        \shortstack{\textbf{Power} \\ \textbf{[kW]}} & 
        \shortstack{\textbf{Purification} \\ \textbf{[kg/h]}} &
        \shortstack{\textbf{with reflux} \\ \textbf{[kg/h]}} &
        \shortstack{\textbf{Power} \\ \textbf{[kW]}} \\
        \midrule
         \textbf{Nominal} & 1200 & 1800 & 46 & 1600 & 2400 & 61 \\
         \textbf{Conservative} & 1800 & 2700 & 69 & 2400 & 3600 & 92
    \end{tabular}
\end{table*}
We assume that the additional gaseous flow required for the reduction of the type 1b sources will scale by detector surface (e.g. number of sensor cables), and is still small compared to the required liquid flows.
Thus, the power estimate only accounts for the required liquid flow to remove type 1a sources.  
The required cooling and heating power was estimated by using the specific enthalpy change of $\Delta h_\mathrm{Xe}(p=\qty{2.2}{\bar})=\qty{92}{\kilo \joule \per \kilo \gram}$ required to fully change the phase between liquid and gaseous xenon \cite{coolprop}. 
The estimated heating and cooling power also account for the successfully tested and required column reflux-ratio $R=0.5$, which means that an additional \qty{50}{\percent} of the total flow are circulated internally inside the column to ensure a stable distillation. Thus, the required heating and cooling power were estimated as 
\begin{equation}
    \dot{Q}^\mathrm{XLZD}_\mathrm{E,i} = (1 + R) \cdot \lambda_\mathrm{dist} \cdot m^\mathrm{XLZD}_\mathrm{i} \cdot \Delta h_\mathrm{Xe}\,,
\end{equation}
not accounting for any additional thermal losses arising from non-perfect insulation, liquid transport, or losses caused across heat exchangers between the distillation column and the heat pump. 
Although highly efficient heat exchanger solutions for the current generation of radon removal systems have already been developed \cite{heat_exchanger}. 
It should also be noted that an auxiliary cooling infrastructure will be required to compensate for these additional losses as well as during the initial cooldown of the system, as the presented heat pump concept can only provide a combined cooling and heating power.
To address these additional questions, more detailed engineering studies are required, but are beyond the scope of this manuscript.

Based on the measurements performed with the presented small proof-of-concept demonstrator, an XLZD sized heat pump providing \qty{60}{\kilo \watt} cooling and \qty{60}{\kilo \watt} heating power, sufficient to cover about three out of the four scenarios, would require 
\begin{equation}
    P^\mathrm{XLZD}_\mathrm{el} = \frac{\dot{Q}_\mathrm{E}}{COP^\mathrm{c}_\mathrm{real}}=\qty{200}{\kilo \watt}
\end{equation}
electrical power.
This is still too high for any future radon removal system as this would put tight constraints on the available lab sides. 
However, it should be noted that this demonstrator was designed as a proof-of-con\-cept and was not optimized for its efficiency. 
The greatest improvement can be achieved by utilizing more efficient gas compressor with a higher $\eta_\mathrm{C}$. 
A new 25-times larger system is currently being constructed within the ERC AdG project ``LowRad''. 
This new system is designed to achieve a xenon mass flow of about \qty{103}{\kilo \gram \per \hour}, while consuming \qty{4}{\kilo \watt} of electrical power. This improves $\eta_\mathrm{C}$ by a factor 1.6 reducing the required power to about \qty{125}{\kilo \watt}. 
Furthermore, we expect, based on past experience with our currently operating krypton and radon distillation systems \cite{xenonnt_kr_removal,radon_removal_system} that other efficiency loss terms like $\eta_\mathrm{B}={\EtaBypass}$, $\eta_\mathrm{T}={\EtaThermal}$, $\eta_\mathrm{P,HE}={\EtaHeatExchanger}$ will not scale proportionally with the mass flow and reduce in larger applications. 
We also aim to optimize and reduce the efficiency loss $\eta_\mathrm{P}={\EtaPressure}$, which is induced by the higher required compressor outlet pressure, but this might be difficult. 
Eventually, the precise design of the radon removal system for XLZD has not been decided yet. 
We expect that the final system will be distributed over three to four distillation systems, enabling maintenance in parallel while taking data with the experiment, and a potential upscaling in case higher purification flows are required.

Comparing the required electrical power of \qty{125}{\kilo \watt} and the provided cooling power of \qty{60}{\kilo \watt} for the proposed heat pump with other commercial solutions indicates that the heat pump concept will be more efficient. 
Especially, the large cooling powers are a challenge as, e.g., LN$_2$ cooling utilizing external LN$_2$ supplies from highly efficient LN$_2$ plants seem unfeasible\footnote{A cooling power of \qty{30}{\kilo \watt} would consume about \qty{13}{tonne} of LN$_2$ per day which is not only a logistical problem as this corresponds to about one full LN$_2$ truck every three days, but also poses a potential hazardous risk when working in an underground facility which must be supplied with breathing air.}.
Other commonly used cooling systems in the required temperature and cooling power range are Turbo-Brayton, Stirling or Gifford-McMahon (GM) cryocoolers. 
For example, the LZ collaboration utilizes a closed loop LN$_2$ Stirling cryocooler, which provides about \qty{1}{\kilo \watt}, while consuming \qty{11}{\kilo \watt} of electrical power \cite{LZ_design_book}. 
A large GM cryocooler of the type AL600 from Cryomech provides about \qty{700}{\watt} cooling power for \qty{12.5}{\kilo \watt} \cite{cryocooler_gm}, while a large Turbo-Brayton based cryocooler of the type TBF-175 of the company AirLiquide can provide on the order of \qty{50}{\kilo \watt} cooling power while consuming \qty{195}{\kilo \watt} of electrical power \cite{cryocooler_tb}. 
None of these numbers include the additional required \qty{60}{\kilo \watt} heating power, which is also already included within the \qty{125}{\kilo \watt} electrical power cost for the proposed heat pump design.

\section{Conclusion}
\label{sec:conclusion}
In this manuscript, we present a first proof-of-concept of a fully hermetically separated small-scale cryogenic heat pump demonstrator.
The heat pump is based on a left-turning Clausius-Rankine cycle using xenon as the working medium.
Two measurements were conducted at nominal pressures of \qty{3.3}{\bar} and \qty{4.3}{\bar}, to test the potential impact of a \qty{1}{\bar} hydrostatic pressure loss in a \qty{3}{\meter} distillation column.
In both scenarios, a cooling and heating power of {\CoolingPower} and {\HeatingPower} were measured, while consuming {\ElectricalPowerThreeBar} and {\ElectricalPowerFourBar} of electrical power, respectively. 
A new 25-times larger heat pump prototype is currently under construction. 
The prototype will be integrated into a fully operational XENONnT-sized radon distillation column to investigate its performance, thermal losses and test its control when coupled to another externally regulated system. 
These new studies will be a stepping stone towards a final XLZD sized system, which requires another order of magnitude scaling to achieve the required cooling and heating power of about \qty{60}{\kilo \watt} each for a purification mass flow of \qty{1600}{\kilo \gram \per \hour}.

\begin{acknowledgements}
We gratefully acknowledge support from the European Research Council (ERC) through the ERC AdG project ``LowRad''.
This project has received funding from the European Research Council under the European Union’s Horizon 2020 research and innovation program (Grant agreement No. 101055063).
% -----
We further, thank the University of M\"unster, the Center for Information Technology (CIT) of the University M\"unster, and the mechanical and electrical workshop of the institute for nuclear physics for hosting and supporting the project. 
\end{acknowledgements}

% BibTeX users please use one of
%\bibliographystyle{spbasic}      % basic style, author-year citations
%\bibliographystyle{spmpsci}      % mathematics and physical sciences
%\bibliographystyle{spphys}       % APS-like style for physics

% Custom bib for arXiv only. For EPJC use one of above.
\bibliographystyle{utphys}
\bibliography{bibliography.bib}   % name your BibTeX data base

\end{document}